\RequirePackage{lineno} 
\documentclass[12pt,onecolumn,prc,showpacs,preprintnumber,amsmath,
floatfix,amssymb]{revtex4-1}

\usepackage{epsfig}
\usepackage{graphicx}
\usepackage{dcolumn}
\usepackage{bm}
\usepackage{multirow}           

\def\var{\mbox{\boldmath $\varepsilon$}}
\def\p{\mbox{\boldmath $p$}}
\def\q{\mbox{\boldmath $q$}}
\def\k{\mbox{\boldmath $k$}}

\allowdisplaybreaks[4]  

\begin{document}

\title{Assessing the accuracy of the GENIE event generator with electron
scattering data: Reduced cross section and nuclear transparency}
\author{A.~V.~Butkevich and S.~V.~Luchuk}
\affiliation{Institute for Nuclear Research,
Russian Academy of Sciences, Moscow 117312, Russia\\}
\date{\today}

\begin{abstract}

  The reduced cross sections of the semi-exclusive $(l,l'p)$ lepton scattering
  process can be identified with distorted nuclear spectral functions.
  Irrespective of the type of interaction the distorted spectral function is
  determined mainly by intrinsic properties of the target and the ejected
  nucleon interaction with residual nucleus. Thus the reduced cross sections
  of the neutrino and electron scattering on nuclei as functions of nucleon
  missing momentum and energy are similar up to Coulomb corrections. Here we
  demonstrate the utility of this approach by benchmarking the GENIE neutrino
  event generator against a broad set of semi-exclusive electron scattering
  data for carbon target. We observe persistent disagreements between the
  generator predictions and data in the range of nucleon missing momenta less
  than 80 MeV/c. The approach presented in this paper provide a great
  opportunity to test better the accuracy of nuclear models of quasi-elastic
  neutrino-nucleus scattering, employed in neutrino event generators.
  
\end{abstract}

\maketitle

\section{Introduction}

The primary physics goal for current~\cite{NOvA1, T2K} and future~\cite{ DUNE,
  HK2T, SBN} accelerator-based neutrino experiments are measuring the lepton
CP violation phase, determining neutrino mass ordering, and testing the
three-flavor paradigm. In these experiments to evaluate the oscillation
parameters, the probabilities of neutrino oscillations as functions of neutrino
energy are measured. The accuracy to which neutrino oscillation parameters can
be extracted depends on the ability of experiments to determine the individual
energy of detected neutrino.

The neutrino beams are not monoenergetic and have broad
distributions that range from tens of MeVs to a few GeVs. At the GeV-
scale neutrino energies the neutrino can interact with a nucleus through a wide
range of reaction channels. These include the charged-current (CC) quasielastic
 (QE) scattering, two-body meson exchange current (MEC) channels, resonance
production and deep inelastic scattering.
The incident neutrino energy is reconstructed using kinematic or
calorimetric methods. At energy about 1 GeV, where the CCQE scattering is
dominant, the incoming neutrino energy can be derived from lepton kinematics
alone. Conservation of total energy in neutrino interactions imply that
$\var_{\nu}=\var_l + \var_h$, where $\var_{\nu}$, $\var_l$, and $\var_h$ are
neutrino, leptonic, and hadronic energies, respectively. The total hadronic
energy deposit is obtained by summing all the visible energy not associated
with lepton. A model dependent fit $\var_h=f(E_{vis})$ obtained from simulation
is used to relate the visible energy $E_{vis}$ to the estimated total hadronic
energy $\var_h$. Thus, the calorimetric method relies not only on the visible
energy measured in the detector, but also on the models of the neutrino-nucleus
interactions that are implemented in neutrino event generators. In addition the
neutrino-nucleus scattering model is critical for obtaining background
estimates, in analysis aimed at determining the neutrino oscillation parameters.

Unfortunately, due to wide range of neutrino energy beams and poor statistics
available from current neutrino experiments, it is challenging to provide
reliable and consistent predictions for the diversity of processes that can take
place in the energy range covered by these neutrino beams. Various contributions
to the cross sections can significantly overlap with each other making it
difficult to identify, diagnose and remedy shortcomings of nuclear models.
Moreover, neutrino scattering measurements are typically reported integrated
over a range of energies and angles. Therefore the task to assess the accuracy
of the neutrino event generators is more complicated than simply comparing
generators predictions to neutrino data. It is necessary to come up with
independent validation methods, which would allow specific physical mechanisms
to be tested.

While electron and neutrino interactions are different at the primary vertex,
many underlying physics processes in the nucleus are very similar.
From the nuclear point of view the influence of nuclear medium
effects such as the nuclear ground state and interaction of the outgoing
nucleon with the residual nucleus can be expected to be largely the same for
electron and for neutrino-induced processes. It is possible to exploit this
similarity and use electron scattering data with precise kinematics and
large statistics to test the neutrino energy reconstruction methods and
interaction models.

Electron and neutrino scattering are strongly linked in theory. Therefore any
model of neutrino interactions (vector+axial)
should also be able to reproduce electron (vector) interactions.
The vector part of the electroweak interaction can be inferred directly from
the electron scattering data. A model unable to reproduce electron measurements
cannot be expected to provide accurate prediction for neutrino cross sections.

Recent years have seen a plethora of analyses of electron-scattering data to
test the vector current part of the lepton-nucleus interaction against
existing inclusive electron scattering cross sections for different target
nuclei.
The relativistic distorted wave impulse
approximation (RDWIA), initially designed for description of exclusive
$(e,e'p)$
data~\cite{Pick, Udias, JKelly} and then adopted for neutrino reactions was
successfully tested against inclusive $(e,e')$ data~\cite{Gon3,BAV7}. The
SuSAv2 model exploits the similarities between both interaction types to guide
the description of weak scattering process~\cite{Megias2, Megias3}.
Neutrino event generators GENIE~\cite{Ankow, e4v1, CLAS, Ruso},
NuWro~\cite{NuWro1, NuWro2}, ACHILLES~\cite{ACHI}, and NEUT~\cite{NEUT} have
been extended to electron scattering and their results were compared with
measured inclusive cross sections of electron-nucleus ($eA$) interaction.

The further progress in reducing systematic errors in
neutrino oscillation experiments requires a more extensive use of measurements
of final-states protons. It is needed for neutrino energy reconstruction,
because more exclusive final state measurement allows to better estimate the
neutrino energy and provide information about multinucleon contribution to the
inclusive cross section. Inclusive reactions
are relatively insensitive to the details of the final nuclear states.
Therefore, rather simple models may yield inclusive and total cross sections
that are not very different from cross sections calculated
in the most sophisticated models, but cannot make predictions on both leptons
and hadrons in the final states.

The event generators model final state interactions (FSI) of knocked-out nucleon
with residual nucleon using the intranuclear cascade models. These models should
be able to reproduce nuclear transparency data from electron scattering.
Nuclear transparency is defined as a probability that a knocked-out nucleon
leaves the residual nucleus without re-interaction. The nucleon cascade models,
implemented in the GENIE~\cite{Dytman} and NuWro~\cite{NuWro3} event generators
were tested against the nuclear transparency data. It was shown that these
generators are able to reproduce these data if short-range nucleon-nucleon
(NN) correlations in the ground state of the nucleus are taken into account.

The semi-exclusive $(l,l'p)$ lepton scattering process involves the specific
asymptotic states and allows to test
more in detail the nuclear model.
The reduced cross section, obtained from the measured differential
semi-exclusive electron scattering cross section dividing by the kinematic
factor and the off-shell electron-proton cross section, can be identified as
the distorted spectral function.
Thus, irrespective of the type of interaction
(electromagnetic or weak) the distorted spectral function is determined mainly
by the intrinsic properties of the target and the ejected nucleon interaction
with residual nucleus and depends upon the initial and ejectile nucleon's
momenta and angle between them. Microscopic and unfactorized RDWIA was used~
\cite{BAV1, BAV2, BAV3, BAV4} for calculation of the charged-current (CC)
quasielastic (QE) neutrino scattering reduced cross sections as function of the
missing nucleon momentum. The results were compared with ones obtained from
measurements of $(e,e^{\prime}p)$ scattering on oxygen, carbon, calcium, and
argon targets. It was shown, that these cross sections are similar to those of
electron scattering data and are in agreement with electron scattering data.
This approach provides novel constraints on nuclear
models of the CCQE scattering and can be applied to test spectral functions and
FSI employed in neutrino event generators.   

The goal of the present paper is to carry out a systematic comparison of the
CCQE reduced cross sections and nuclear transparency calculated within the
models employed in the GENIE event generator~\cite{GEN1, GEN2} with ones
measured in electron scattering on carbon targets. For carbon targets there are
high-precision data sets with different monochromatic electron energies and
scattering angles. We consider GENIE because this generator is at present
employed by all Femilab-based neutrino experiments.

This paper is organized as follows. In Sec.II we present briefly the formalism
needed to describe  the semi-exclusive lepton-nucleus CCQE scattering process.
Reduced cross sections and nuclear transparency are defined.
Calculations of the reduced cross sections and nuclear transparency with
neutrino event generator are described  in Sec.III. In Sec.IV we regard the
comprehensive model configurations employed in the GENIE v.3 simulation
framework.  
Results and discussion of the calculations are presented in Sec.IV. Our
conclusions are summarized in Sec.V.

\section{Formalism of quasi-elastic scattering: reduced cross section and
  nuclear transparency}

In the laboratory frame the differential cross section for exclusive
electron ($\sigma ^{el}$) and (anti)neutrino ($\sigma ^{cc}$) CC scattering can
be written as
\begin{subequations}
\begin{align}
\label{1} 
\frac{d^6\sigma^{el}}{d\varepsilon_f d\Omega_f d\varepsilon_x d\Omega_x} 
&=
\frac{\vert\p_x\vert\varepsilon_x}{(2\pi)^3}\frac{\varepsilon_f}
{\varepsilon_i}
 \frac{\alpha^2}{Q^4} L_{\mu \nu}^{(el)}\mathcal{W}^{\mu \nu (el)}
\\                                                                  
\frac{d^6\sigma^{cc}}{d\varepsilon_f d\Omega_f d\varepsilon_x d\Omega_x} 
&=
\label{2} 
\frac{\vert\p_x\vert\varepsilon_x}{(2\pi)^5}\frac{\vert\k_f\vert}
{\varepsilon_i} \frac{G^2\cos^2\theta_c}{2} L_{\mu \nu}^{(cc)}
\mathcal{W}^{\mu \nu (cc)},
\end{align}
\end{subequations}
where
$k_i=(\varepsilon_i,\k_i)$ and $k_f=(\varepsilon_f,\k_f)$ are initial and
final lepton momenta, $p_A=(\varepsilon_A,\p_A)$, and
$p_B=(\varepsilon_B,\p_B)$ are the initial and final target momenta,
$p_x=(\varepsilon_x,\p_x)$ is the ejectile nucleon momentum, $q=(\omega,\q)$ 
is the momentum transfer carried by the virtual photon (W-boson), and
$Q^2=-q^2=\q^2-\omega^2$ is the photon (W-boson) virtuality. In
Eqs. (\ref{1},\ref{2}) $\Omega_f$ is the solid angle for the lepton momentum,
$\Omega_x$ is the
 solid angle for the ejectile nucleon momentum, $\alpha\simeq 1/137$ is the
fine-structure constant, $G \simeq$ 1.16639 $\times 10^{-11}$ MeV$^{-2}$ is
the Fermi constant, $\theta_C$ is the Cabbibo angle
($\cos \theta_C \approx$ 0.9749), $L^{\mu \nu}$ is the lepton tensor, and
 $\mathcal{W}^{(el)}_{\mu \nu}$ and $\mathcal{W}^{(cc)}_{\mu \nu}$ are
correspondingly the electromagnetic and weak CC nuclear tensors.

For exclusive reactions in which only a single discrete state or a narrow
resonance of the target is excited, it is possible to integrate over the
peak in missing energy and obtain a fivefold differential cross section of
the form
\begin{subequations}
\begin{align}
\label{cs5:el}
\frac{d^5\sigma^{el}}{d\varepsilon_f d\Omega_f d\Omega_x} &= R
\frac{\vert\p_x\vert\tilde{\varepsilon}_x}{(2\pi)^3}\frac{\varepsilon_f}
{\varepsilon_i} \frac{\alpha^2}{Q^4} L_{\mu \nu}^{(el)}W^{\mu \nu (el)}
\\                                                                       
\label{cs5:cc}
\frac{d^5\sigma^{cc}}{d\varepsilon_f d\Omega_f d\Omega_x} &= R
\frac{\vert\p_x\vert\tilde{\varepsilon}_x}{(2\pi)^5}\frac{\vert\k_f\vert}
{\varepsilon_i} \frac{G^2\cos^2\theta_c}{2} L_{\mu \nu}^{(cc)}W^{\mu \nu (cc)},
\end{align}
\end{subequations}
where $R$ is a recoil factor
  \begin{equation}\label{Rec}
R =\int d\varepsilon_x \delta(\varepsilon_x + \varepsilon_B - \omega -m_A)=
{\bigg\vert 1- \frac{\tilde{\varepsilon}_x}{\varepsilon_B}
\frac{\p_x\cdot \p_B}{\p_x\cdot \p_x}\bigg\vert}^{-1},                    
\end{equation}
$\tilde{\varepsilon}_x$ is the solution to the equation
$
\varepsilon_x+\varepsilon_B-m_A-\omega=0,
$
where $\varepsilon_B=\sqrt{m^2_B+\p^2_B}$, $~\p_B=\q-\p_x$ and $m_A$ and $m_B$
are masses of the target and recoil nucleus, respectively. Note that missing
momentum is $\p_m=\p_x-\q$ and missing energy $\var_m$ are defined by
$\var_m=m + m_B -m_A$.
The invariant electromagnetic (weak CC) matrix element is represented by the
contraction of electron (neutrino) and nuclear response tensors of the form
 \begin{equation}
 \label{avsec}   
L_{\mu \nu}^{(el)(cc)} = \langle j_{\mu}^{(el)(cc)}j_{\nu}^{+ (el)(cc)}\rangle     
\end{equation}
 \begin{equation}
 \label{avsec}   
W_{\mu \nu}^{(el)(cc)} = \langle J_{\mu}^{(el)(cc)}J_{\nu}^{+ (el)(cc)}\rangle,   
\end{equation}
 where $j^{\mu}$ is the electron or neutrino current, $J_{\mu}^{(el)(cc)}$ is a
 matrix element of the nuclear electromagnetic or CC current, and the
 angled brackets denote averages over initial states and sums over final states.

 The reduced cross section is given by 
\begin{equation}
\sigma_{red} = \frac{d^5\sigma^{(el)(cc)}}{d\varepsilon_f d\Omega_f d\Omega_x}
/K^{(el)(cc)}\sigma_{lN},                                                 
\end{equation}
where
$K^{el} = R {p_x\varepsilon_x}/{(2\pi)^3}$ and
$K^{cc}=R {p_x\varepsilon_x}/{(2\pi)^5}$
are phase-space factors for electron and neutrino scattering  and 
 \begin{equation}
 \sigma_{eN}^{(el)} = \frac{\var_f}{\var_i}\frac{\alpha^2}{Q^4}(L^{(el)}_{\mu \nu}
 W^{\mu \nu (el)})_{PWIA}                                           
\end{equation}
 \begin{equation}
   \sigma_{\nu N}^{(cc)} = \frac{k_f}{\var_i}\frac{(G_F \cos\theta_c)^2}{(2\pi)^2}
   (L^{(cc)}_{\mu \nu}W^{\mu \nu (cc)})_{PWIA}                         
\end{equation}
are the corresponding
elementary cross sections for the electron $\sigma_{eN}$ and neutrino
$\sigma_{\nu N}$ scattering on the moving free nucleon in the plane wave
impulse approximation (PWIA) normalized to
unit flux. The PWIA response tensor is computed for a free nucleon in the
final state and is given by 
 \begin{equation}
   W_{PWIA}^{\mu \nu} = \frac{1}{2}{\rm Trace} (J^{\mu}J^{\nu +}),           
\end{equation}
 where
 \begin{equation}
 J^{\mu}_{s_j,s_i}=\sqrt{\frac{m^2}{\var_i \var_j}}\bar{u}({\p_f},s_j)  
 \Gamma^{\mu}u({\p_i},s_i)
\end{equation}
 is the single-nucleon current between free spinors normalized to unit flux. The
 initial momentum ${\p}_i = {\p}_f - {\q}_{eff}$ is obtained from the
 final ejectile momentum and the effective momentum transfer ${\q}_{eff}$ in
 laboratory frame, and the initial energy is placed on shell. The effective
 momentum transfer accounts for incoming electron acceleration in the nuclear
 Coulomb field and ${\q}_{eff} = {\q}$ for the incoming neutrino.

 The single-nucleon charged current  has the $V{-}A$
 structure $J^{\mu (cc)} = J^{\mu}_V + J^{\mu}_A$. For calculation of vertex function
 $\Gamma^{\mu (cc)} = \Gamma^{\mu}_V + \Gamma^{\mu}_A$ of a moving but free nucleon
 it is simpler to use the CC1 prescription for vector current vertex function
\begin{equation}
\Gamma^{\mu}_V = G_M(Q^2)\gamma^{\mu} - \frac{{\bar P}^{\mu}}{2 m}F_M(Q^2)  
\end{equation}
and the axial current vertex function
\begin{equation} 
\Gamma^{\mu}_A = F_A(Q^2)\gamma^{\mu}\gamma_5 + F_P(Q^2)q^{\mu}\gamma_5,  
\end{equation}
where $\bar{P}=(\var_f + \bar{\var}, 2\p_x - \q)$ and
$ \bar{\var}=\sqrt{m^2 + (\p_x -\q)^2}$.
The weak vector form factors $F_V$ and $F_M$ are related to corresponding
electromagnetic ones for proton $F^{(el)}_{i,p}$ and neutron $F^{(el)}_{i,n}$
by the hypothesis of conserved vector current (CVC)
\begin{equation}
F_i = F^{(el)}_{i,p} - F^{(el)}_{i,n}~~~~(i=V,M).                          
\end{equation}
For the axial $F_A$ and pseudoscalar $F_P$ form factors we use the dipole
approximation.
 
The reduced cross section can be  regarded as the nucleon momentum distribution
modified by FSI, i.e. as the distorted spectral function. Final-state
interactions make the reduced cross sections $\sigma_{red}({\p}_m, {\p}_x)$
depend upon ejectile momentum ${\p}_x$, angle between the initial and final
nucleon momentum and upon incident lepton energy. In the nonrelativistic PWIA
limit, $\sigma_{red}$ reduces to the bound-nucleon momentum distribution.
These cross sections for (anti)neutrino  scattering off
nuclei are similar to the electron scattering apart from small differences at
low beam energy due to effects of Coulomb distortion of the incoming electron
wave function as shown in Refs.~\cite{BAV1, BAV2, BAV3}.  

The factorization approximation to the knockout cross section stipulates that
\begin{equation}
\frac{d^5\sigma^{(el)(cc)}}{d\varepsilon_f d\Omega_f d\Omega_x}=
K^{(el)(cc)}\times \sigma_{lN} \times  \sigma_{red}(\var_m, {\p}_m, {\p}_x).    
\end{equation}
This factorization implies that the initial nuclear state and FSI effects are 
decoupled from the leptonic vertex with preserved  correlations between the
final lepton and nucleon. In the nonrelativistic PWIA limit the fivefold
differential cross section Eq.(\ref{1}) and Eq.(\ref{2}) may be expressed as
product of the
phase-space factor K, elementary cross section $\sigma_{eN}$, and nucleon
momentum distribution $S(\p_m,\var_m)$. Such factorization is not strictly
valid relativistically because the binding potential alters the relationship
between lower and upper components of a Dirac wave function~\cite{Caballero}.

The nuclear transparency measured as a function of $Q^2$, is defined as
\begin{equation}
  \label{trans}
  T_{exp}(Q^2)=\frac{\int_Vd^3p_md\var_mY_{exp}(\var_m,\p_m)}
  {f\int_Vd^3p_md\var_mY_{PWIA}(\var_m,\p_m)},    
\end{equation}
where $Y_{exp}$ and $Y_{PWIA}$ are proton yields and theoretical calculations
within the PWIA, respectively and $f$ is a correction factor describing the
depletion of the single-hole spectral function. The PWIA model describe the
nuclear structure, assuming the independ particle shell model (IPSM). The
calculation within the IPSM overestimate single-particle strength, because the
shells are not fully occupied due to the short-range nucleon-nucleon (NN)
correlations, leading to the
appearance of high-energy and high-momentum components in the ground state of
target. The phase space volume $V$ is restricted to the quasielastic region by
the conditions $\var_m \le 80$ MeV and $|\p_m| \le 300$ MeV/c. One should note
that measured nuclear transparency is model dependent, because it relies on the
accuracy of PWIA calculations.

Similarly, a theoretical definition of nuclear transparency takes the form
\begin{equation}
\label{tran} 
  T_{th}(Q^2)=\frac{\sum_{\alpha}\int dp_m p_m\sigma_{FSI}(\var_{\alpha},\p_m)}
  {\sum_{\alpha}\int dp_m p_m\sigma_{PWIA}(\var_{\alpha},\p_m)},    
\end{equation}
where the numerator integrates the reduced cross section calculated within
nuclear model, that includes the distortion due to FSI and the denominator
integrates the PWIA cross section for the same model of the spectral function.
In Eq.~(\ref{tran}) sums are taken over orbits $\alpha$ and $\var_{\alpha}$ is the
missing energy for orbital $\alpha$.

\section{Neutrino event generator as a tool in reduced cross section and
transparency studies}

To test nuclear models employed in the GENIE event generator we compare
reduced cross sections and nuclear transparency data for carbon nucleus with
predictions made within GENIE simulation framework. The carbon nucleus which is
by far most extensively studied in electron scattering experiments was used.
This framework offers several models for nuclear ground state, several models
for each of the $eA$ or $\nu A$ scattering mechanisms and several models for
hadronic final interaction, i.e. intranuclear rescattering of the outgoing
hadrons. 

Only CCQE neutrino interactions with carbon nucleus were simulated. To select
events and their kinematics in the stable final state which correspond to a
$1\mu1p + X$ signal, we choice events with one proton and any number of
neutrons. We then make sampling of $1\mu1p$ events from cascade model which
have not been affected by inelastic FSI. Two methods are used for this.
The first based on the topology of the events in the GENIE cascade, and
second based solely on the knowledge of the kinematics of events
Ref.~\cite{Sanchez}.

For the first the GENIE output which yield for every event a number of hadrons
emitted from the nucleus was used. When only
one proton track is presented in an event, the original proton escapes the
nucleus without inelastic interactions. When multiple tracks are present,
FSI has occurred and the original proton generally change energy. On the
other hand, one may want to make a selection of events based purely on the
lepton and nucleon kinematics. As we are simulating, we know the true incoming
energy and we can define for each event a missing energy as 
 \begin{equation}
 \tilde{\var}_m = \var_{\nu} - \var_{\mu} -T_p = \omega - T_p.    
\end{equation}
 This new definition of the missing energy does not take into account explicitly
the recoil energy of the residual hadronic system. In electron scattering
experiments where the initial energy is known the events with
$\var_{min} \le \tilde{\var}_m \le \var_{max}$ are separated.

We make a selection of events,
after applying the GENIE cascade, in which only those events that corresponds
to the range of $\tilde{\var}_m \le 80$ MeV are present.
In ``perpendicular kinematics'' used in electron scattering experiments
Refs.~\cite{Saclay, Tokyo1, Kelly2, SLAC, JLab} the incident energy and
energy transfered are fixed,
as are the electron scattering angle and the outgoing proton energy, whereas
the missing momentum changes with the proton angle. It is necessary to choose
the electron angle and outgoing proton energy such that $|\p_x|=|\q|$. The
vectors $\p_m$ and $\p_x$ are almost perpendicular.   

As in all electron scattering experiments the lepton and hadron spectrometers
are set in plane, the out-of-plane angles are fixed to $\phi_e=0^{\circ}$ and
$\phi_p=180^{\circ}$. The sixfold exclusive cross section is given by
\begin{equation}
  \frac{d^6\sigma}{d\var_{\mu} d\Omega_{\mu} dT_p d \Omega_p} =
  \frac{N_{1\mu 1p}}{N_{tot}}\frac{\sigma_{tot}(\var_{\nu})}
       {\Delta \var_{\mu} \Delta T_p \Delta \Omega_{\mu} \Delta \Omega_p},  
\end{equation}
where $N_{tot}$ is the total number of generated CCQE neutrino events with a
total CCQE cross section $\sigma_{tot}(\var_{\nu})$ at energy $\var_{\nu}$.
$N_{1\mu 1p}$ is the number of selected $1\mu 1p$ events in the differential
phase space volume bin
$\Delta V = \Delta \var_{\mu} \Delta T_p \Delta \Omega_{\mu} \Delta \Omega_p$ with
the central values
$\var_{\mu},T_p,\Omega_{\mu}, \Omega_p$ and with size of differential bins 
 $\Delta \var_{\mu}, \Delta T_p, \Delta \theta_{\mu}, \Delta \theta_p,
\Delta \phi_{\mu}, \Delta \phi_p$.
The central values and size of the differential bins of kinematic variables
are given in Table I and Table II, corresponding, for every data set that is
analyzed in this work.
\begin{table}[t]
        \def\arraystretch{1}
         \begin{tabular}{|c| c c c c c c c |}

                \hline \hline
 \multirow{1}{*}{data set} & $\var_{i}$ & $\var_{f}$ & $\theta_{e}$ & T$_{p}$ &
  $\theta_{p}$ & $\Delta\var_{m}$ & Q$^{2}$\\ 
                & (MeV) & (MeV) & (deg) & (MeV) & (deg) & (MeV) & (GeV$^{2}$)\\
                \hline \hline
   \multirow{1}{*}{Tokyo \cite{Tokyo1,Kelly2}} & \multirow{1}{*}{700} &
   \multirow{1}{*}{525} & \multirow{1}{*}{52.3} & \multirow{1}{*}{156} &
   \multirow{1}{*}{49.7 - 70.2} & \multirow{1}{*}{80} & \multirow{1}{*}{0.29} \\
   \multirow{1}{*}{Saclay \cite{Saclay}} & \multirow{1}{*}{497} &
   \multirow{1}{*}{375} & \multirow{1}{*}{52.9} & \multirow{1}{*}{87} &
   \multirow{1}{*}{48 - 91} & \multirow{1}{*}{60} &\multirow{1}{*}{0.16}\\  
   \multirow{1}{*}{SLAC \cite{SLAC}} & \multirow{1}{*}{2015} &
   \multirow{1}{*}{1390} & \multirow{1}{*}{35.5} & \multirow{1}{*}{600} &
   \multirow{1}{*}{43.4 - 54.6} & \multirow{1}{*}{80} & \multirow{1}{*}{1.11} \\
   \multirow{1}{*}{JLab \cite{JLab}} & \multirow{1}{*}{2445} &
   \multirow{1}{*}{2075} & \multirow{1}{*}{20.5} & \multirow{1}{*}{350} &
   \multirow{1}{*}{35.4 - 75.4} & \multirow{1}{*}{80} & \multirow{1}{*}{0.64} \\
                \hline \hline
        \end{tabular}
        \def\arraystretch{1.0}
        \caption{Summary of data for $^{12}$C$(e,e^{\prime}p)$. $\var_{i}$
        is the beam energy, $\var_{f}$ is the central electron energy,
          $\theta_{e}$ is the central electron angle, T$_{p}$ is the central
          proton kinetic energy, $\theta_{p}$ is the central proton angle,
          $\Delta\var_m$ is the range of the missing energy.}
        \label{tab:kinematics1}
\end{table}
\begin{table}[t]
        \def\arraystretch{1.0}
        \begin{tabular}{|c|c c c c c c|}

                \hline\hline
                \multirow{2}{*}{data set} & $\Delta\varepsilon_{f}$ &
                $\Delta\theta_{e}$ & $\Delta$T$_{p}$ & $\Delta\theta_{p}$ &
                $\Delta\phi_{e}$ & $\Delta\phi_{p}$ \\
                & (\%) & (deg) & (\%) & (deg) & (deg) & (deg)\\
                \hline \hline

         \multirow{1}{*}{Tokyo \cite{Tokyo1,Kelly2}} & \multirow{1}{*}{9.5} &
         \multirow{1}{*}{5} & \multirow{1}{*}{38} & \multirow{1}{*}{3} &
         \multirow{1}{*}{3} & \multirow{1}{*}{3} \\

         \multirow{1}{*}{Saclay \cite{Saclay}} & \multirow{1}{*}{18} &
         \multirow{1}{*}{2} & \multirow{1}{*}{23} & \multirow{1}{*}{2} &
         \multirow{1}{*}{6} & \multirow{1}{*}{6} \\

            \multirow{1}{*}{SLAC \cite{SLAC}} & \multirow{1}{*}{4} &
            \multirow{1}{*}{3} & \multirow{1}{*}{8} & \multirow{1}{*}{1} &
            \multirow{1}{*}{3} & \multirow{1}{*}{3} \\

            \multirow{1}{*}{JLab \cite{JLab}} & \multirow{1}{*}{4} &
            \multirow{1}{*}{1.5} & \multirow{1}{*}{14} & \multirow{1}{*}{1} &
            \multirow{1}{*}{3} & \multirow{1}{*}{3} \\
            \hline \hline
            
        \end{tabular}
        \def\arraystretch{1.0}
        \caption{Table of the experimental cuts that used in simulation.
          $\Delta\varepsilon_{f}$ is the electron energy acceptance,
          $\Delta\theta_{e}$ is the electron angle acceptance,
          $\Delta$T$_{p}$ is the proton kinetic energy acceptance, and
          $\Delta\phi_{p}$ is the proton angle acceptance, $\Delta\phi_{e}$ is
          the electron plane angle acceptance.}
        \label{tab:kinematics2}
\end{table}
To go from measured variables $\var_{\mu}$ and $T_p$ to
\begin{equation}
 \var_m = \omega -T_p - \var_B                               
\end{equation}
and
\begin{equation}
  p_m = [p^2_x + k^2_{\nu} + k^2_{\mu} -2k_{\mu} k_{\nu}\cos\theta_{\mu} -
    2p_xk_{\nu}\cos\theta_p + 2k_{\mu}p_x\cos\theta_{\mu p}]^{1/2}        
\end{equation}
we use the formula
\begin{equation}
  \label{pm}
 \frac{d^6\sigma}{dp_m d\var_m d\Omega_{\mu}d \Omega_p}=
  \frac{d^6\sigma}{d\var_{\mu} dT_p d\Omega_{\mu} d \Omega_p}/J(\theta_p),  
\end{equation}
where Jacobin $J=\partial(\var_m,p_m)/\partial(\var_{\mu},T_p)$ is equal to
\begin{equation}
J(\theta_p)=\frac{1}{p_m}(k_{\mu} - k_{\nu}\cos\theta_{\mu} + p_x\cos\theta_{\mu p})
  - \frac{\var_x}{2 p_m}\frac{p^2_x + p^2_m -|\q|^2}{p^2_x}.       
\end{equation}

In Eq.(\ref{pm}) $\cos\theta_{\mu p}=\cos(\theta_{\mu} + \theta_p)$ and we
assess
positive (negative) values to $p_m$ for the condition $\theta_p <\theta_q
(\theta_p > \theta_q)$, where $\cos\theta_q = \p_x\cdot\q/(|\p_x||\q|)$. Then
we can determine
\begin{equation}
 \frac{d^2\sigma}{dp_m d\var_m}=
  \frac{d^6\sigma}{dp_m d\var_m d\Omega_{\mu} d \Omega_p},  
\end{equation}
as a function of missing energy and missing nucleon momentum.
All nuclear models which implemented in the GENIE generator do not allow
calculation of the $d^2\sigma/dp_m d\var_m$ distribution as function of
$\var_m$, as well as the calculation of the $d\sigma/dp_m$ in ranges 
$10 < \var_m <25$ MeV ($1p_{3/2}$ shell) and $ 30< \var_m < 50$ MeV
($1s_{1/2}$ shell), because they use one-dimensional $d\sigma/dp_m$ distributions
with fixed values of binding energy. Therefore we can evaluate only the
$d\sigma/dp_m$ cross section 
\begin{equation}
 \frac{d\sigma}{dp_m}= \Delta \var_m \frac{d^2\sigma}{dp_m d\var_m}  
\end{equation}
as a function of $p_m$, integrated over range $\Delta \var_m$, that is given in
Table I. Then the reduced cross section can be written as
\begin{equation}
  \sigma^{red}(p_m)=\frac{1}{K\sigma_{\nu N}}\frac{d\sigma}{dp_m}(p_m).  
\end{equation}
The calculation of $\sigma_{\nu N}$ is performed with the same nucleon form
factors as GENIE $\sigma/dp_m$ cross section calculation.

The experimental definition of the nuclear transparency Eq.(\ref{trans}), which
is given by ratio of the observed number of $1\mu 1p$ events (without FSI) to
the ones predicted in the PWIA as a function of $T_p$ was used. We take into
account a finite energy acceptance $\Delta T_p/T_p \approx 0.2$ of hadronic
spectrometers.    

\section{Comprehensive model configurations in GENIE framework}

As mentioned above, the GENIE v3 simulation framework uses a few different
models of nucleon momentum distribution in the nuclear ground state. It also
offers several models of quasielastic lepton-nucleus interactions and several
intranuclear cascade models for FSI.

We use the GENIE Version 3.04.00 and consider five distinct sets of GENIEs
models for CCQE scattering namely: the G18-02a (RFG), G18-10a (LFG), G21-11a
(SuSAv2-MEC), sf1d (SF), and effsf (effective spectral function) models
Refs~\cite{G1, G2, G3}. These models are accompanied by the FSI models
available in the GENIE~\cite{G1}. The effective intranuclear model INTRANUKE
hA is an empirical data driven method, uses the cross sections of pions and
nucleons interactions with nuclei as a function of energy up to 1.2 GeV. For
higher energies the GEM03 calculation~\cite{GEM}, normalized to low energy
data is used. The INTRANUKE hN model is a full intranuclear cascade calculation
of pion, kaon, photon, and nucleons interactions with nuclei up to 1.2 GeV.
\begin{figure*}
  \begin{center}
    \includegraphics[height=7cm, width=15cm]{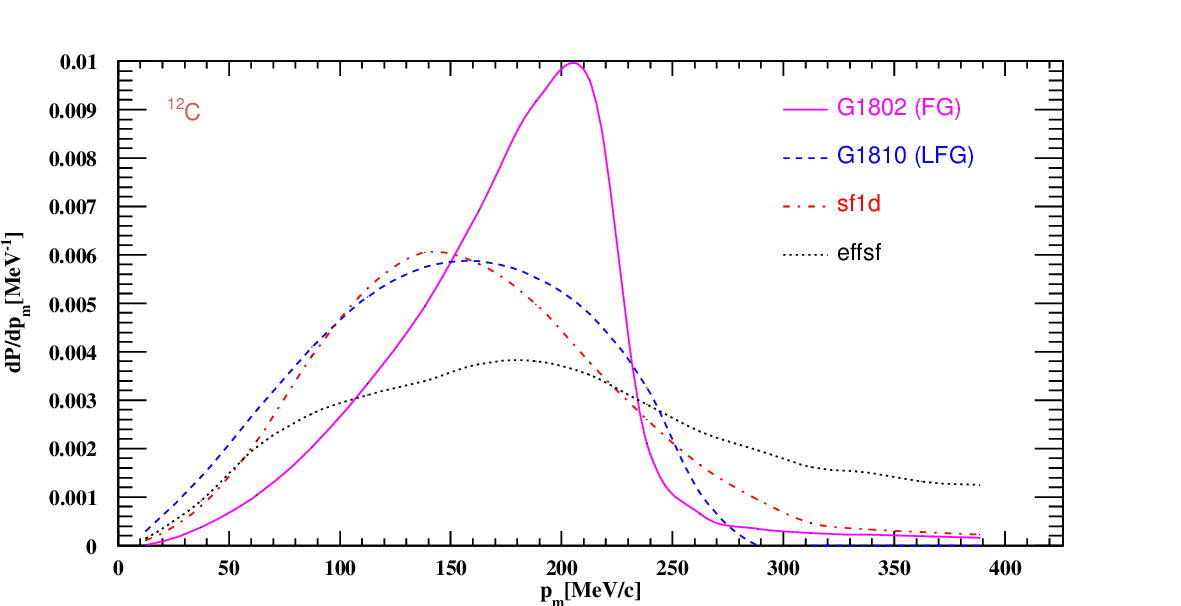}
  \end{center}
  \caption{\label{Fig1}
Initial nucleon momentum distribution for ${}^{12}$C according to GENIE
implementation of G18\_02a (FG, solid line), G18\_10a (LFG, dashed line),
sf1d (dashed-dotted), and effsf (dotted line) models.
}
\end{figure*}

\begin{figure*}
  \begin{center}
    \includegraphics[height=8cm, width=18cm]{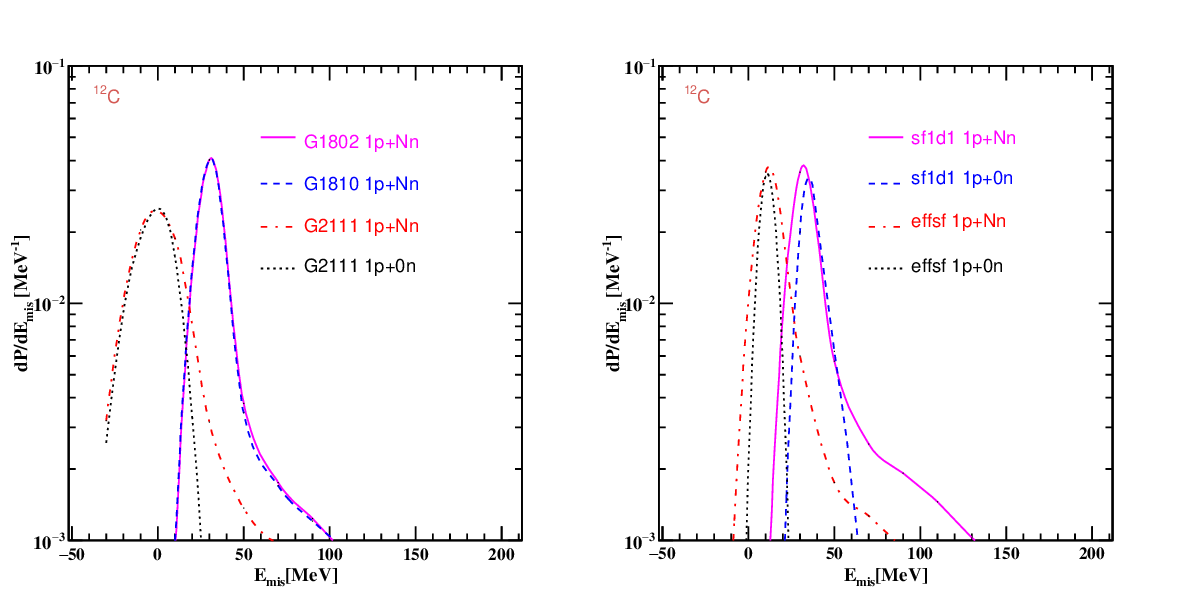}
  \end{center}
  \caption{\label{Fig2}
    Probability density distribution vs missing energy $E_{mis}$ calculated for
    neutrino
energy $\var_{\nu}=0.5$ GeV. Left panel: PDF for (1p+Nn) events calculated with
the G18\_02 (solid line), G18\_10 (dashed line), G21\_11 (dashed-dotted line)
models and for (1p+0n) events calculated within G21\_11 (dotted line) model.
Right panel: PDF for (1p+0n) set calculated with sf1d (solid line), effsh
(dotted line) model and PDF for (1p+Nn) set calculated within the sfd1
(dashed line) and effsf (dashed-dotted line) models. 
}
\end{figure*}

The G18\_02a configuration of GENIE used for CCQE interaction on nuclei 
relies on implementation of the relativistic Fermi gas model (RFG) which has
been modified to incorporate short-range NN-correlations~\cite{BodR}. For
quasielastic and elastic scattering, Pauli blocking is applied and CCQE
scattering is simulated by the Llewellyn-Smith model~\cite{LS}.
The G18\_10a model set includes the full Valencia model~\cite{Val1, Val2, Val3}
for the local Fermi gas (LFG) nucleon momentum distribution. It also uses 
the random phase approximation (RPA) which is a description of long-range and
short-range NN-correlations. The main effect of the RPA corrections is a
suppression of the CCQE cross section at low $Q^2$. 
The G21\_11 model, based on the SuSAv2 approach uses superscaling to write
the inclusive CCQE cross section in terms of a universal function~\cite{Dolan}.
 The superscaling phenomen is shown by electron-nucleus scattering data. The
 nucleon kinematics in this model are determined by choosing its initial
 momentum from the LFG distribution.

The sf1d model that is included in GENIE simulation framework as an
option SpectralFunction1d is based on spectral functions (SF) for finite nuclei
derived from correlated basis function approach~\cite{Benh1, Benh2}. The
SF is a sum of mean field and correlation terms with distinct energy and
momentum distribution. The mean field piece has been fited to the  (e,e'p)
scattering data and the correlation contribution is computed using the local
density approximation. The spectrals function for proton and neutron are the
same. The effective spectral function (effsf) model with or without enhancement
contribution~\cite{Bod1, Bod2} is implemented in the GENIE as an option
EffectiveSF. In this model the magnetic form factor for bound nucleon is
modified to describe the  enhancement of transverse contribution in the
quasielastic cross section of electron-carbon scattering.
In this work all five models are accompanied by the FSI using the effective
intranuclear cascade model INTRANUKE hA.

Figure~\ref{Fig1} displays the bound nucleon momentum distribution in carbon
for the
genuine CCQE events produced using the GENIE and the RFG, LFG, sf1d and effsf
representations of the target nucleus. It is clearly visible that the
probability density function (PDF) $dP/dp_m$ calculated with the RFG in the
range $150 \le p_m \le 220$ ($p_m \le 50$) (MeV/c) is much larger (smaller)
than PDFs used in the other
models, plus correlation tails are observed in the RFG, sf1d, and effsf
distributions. As mentioned before, we consider two sets of events in which the
neutrino knocks out a proton. The first set contains events with a muon and
only one proton (1p + 0n) in the final state. Such events are selected in
electron scattering experiments (e,e'p) to measure reduced cross sections. The
second set contains events with a muon, one proton, and at least one neutron
(1p + Nn) in the final state.
\begin{figure*}
  \begin{center}
    \includegraphics[height=12cm, width=18cm]{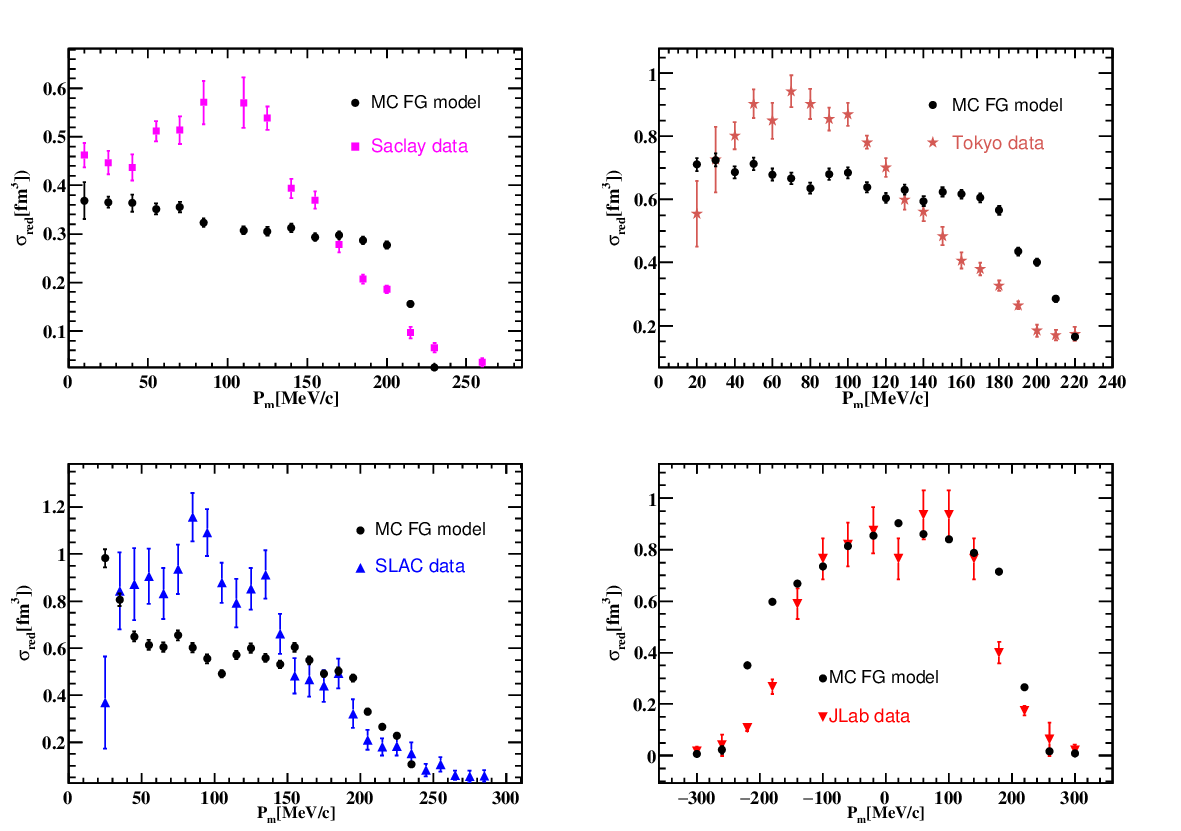}
  \end{center}
  \caption{\label{Fig3}
Comparison of the GENIE G18\_02a (FG model) calculation reduced cross sections
with data as function of missing momentum. As shown in the key, cross sections
were calculated for Saclay~\cite{Saclay}, Tokyo~\cite{Tokyo1, Kelly2},
SLAC~\cite{SLAC}, and JLab~\cite{JLab} kinematics. 
}
\end{figure*}
In neutrino experiments these two sets of events
are not distinguished, because the incident neutrino energy is unknown.
The missing energy $E_{mis}=\var_{\nu}-\var_{\mu}-T_p$ distributions for
(1p+0n) and (1p+Nn) (N$\le$3) sets calculated for neutrino energy
$\var_{\nu} \le 0.5$ GeV using five GENIE' models are shown in Fig.~\ref{Fig2}.
For all models (except G21\_11 model) maximum of distribution is located in the
range $E_m \le 40$ MeV. For the G18\_02a and G18\_10a models the distributions
of events in (1p+0n) set have narrow peak in the range $20 \le E_m \le 40$ MeV.
The inelastic scattering changes the energy of the knocked out protons
significantly and therefore the distributions of events in the (1p+Nn) set are
continued into higher missing region. 
\begin{figure*}
  \begin{center}
    \includegraphics[height=12cm, width=18cm]{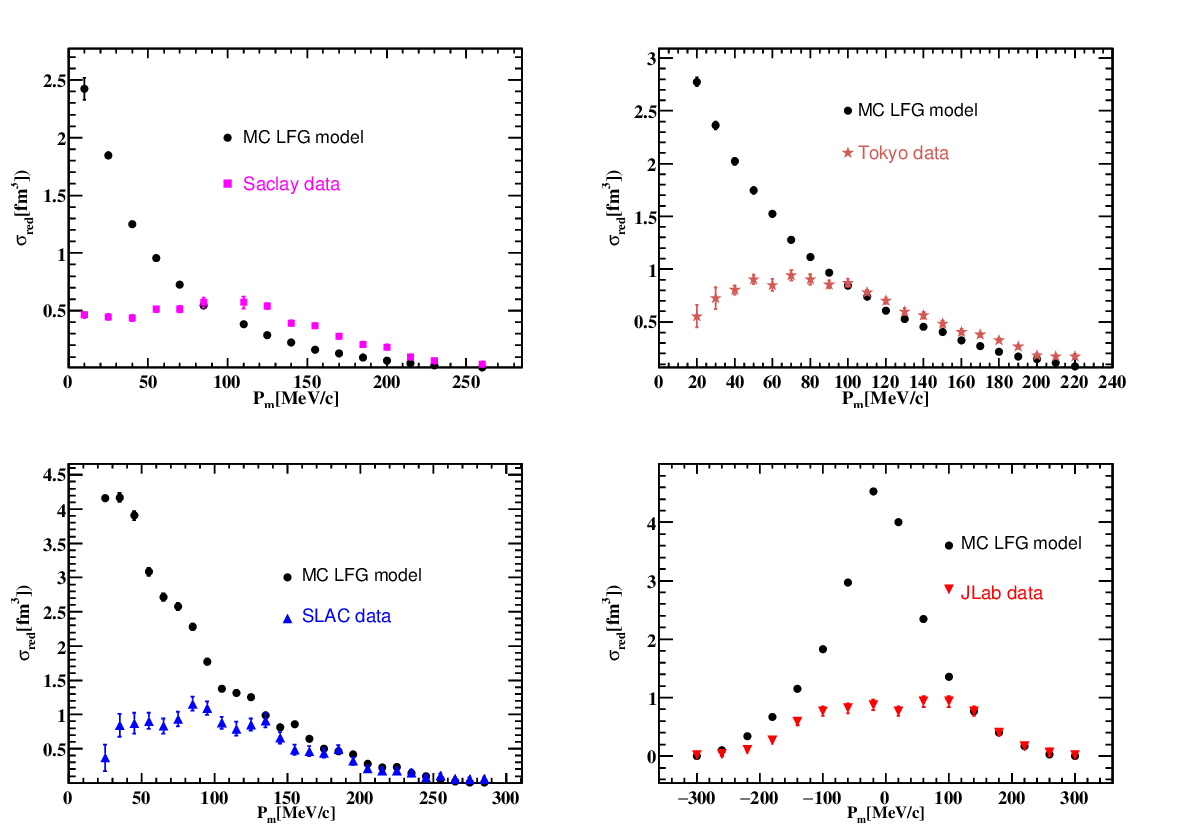}
  \end{center}
  \caption{\label{Fig4}
 Same as Fig.~\ref{Fig3} but for the G18\_10a (LFG) model calculation.    
}
\end{figure*}
\section{Results and analysis}

For each model considered we generated  $10^8$ CCQE neutrino events. 
The reduced cross sections calculated with the GENIE G18\_02a
model that uses the RFG nucleon momentum distribution are shown in
Fig.~\ref{Fig3}
together with Saclay~\cite{Saclay}, Tokyo~\cite{Tokyo1, Kelly2},
SLAC~\cite{SLAC}, and JLab~\cite{JLab} data. It should be noted that negative
values of $p_m$ corresponds to $\phi=\pi$ and positive ones to $\phi=0$. The
cross sections were calculated using the kinematic conditions of data examined
that is presented in Tables I and II.
The values of measured cross sections at the maximum are higher
than calculated ones with exception for JLab kinematics. Discrepancies
up to 50\% at the maximum are observed with low beam energy (low $Q^2$) for
Saclay and Tokyo data and they decrease with the increasing the beam energy.
There is an overall agreement with JLab data at $E_{\nu}=2.4$ GeV.  
\begin{figure*}
  \begin{center}
    \includegraphics[height=12cm, width=18cm]{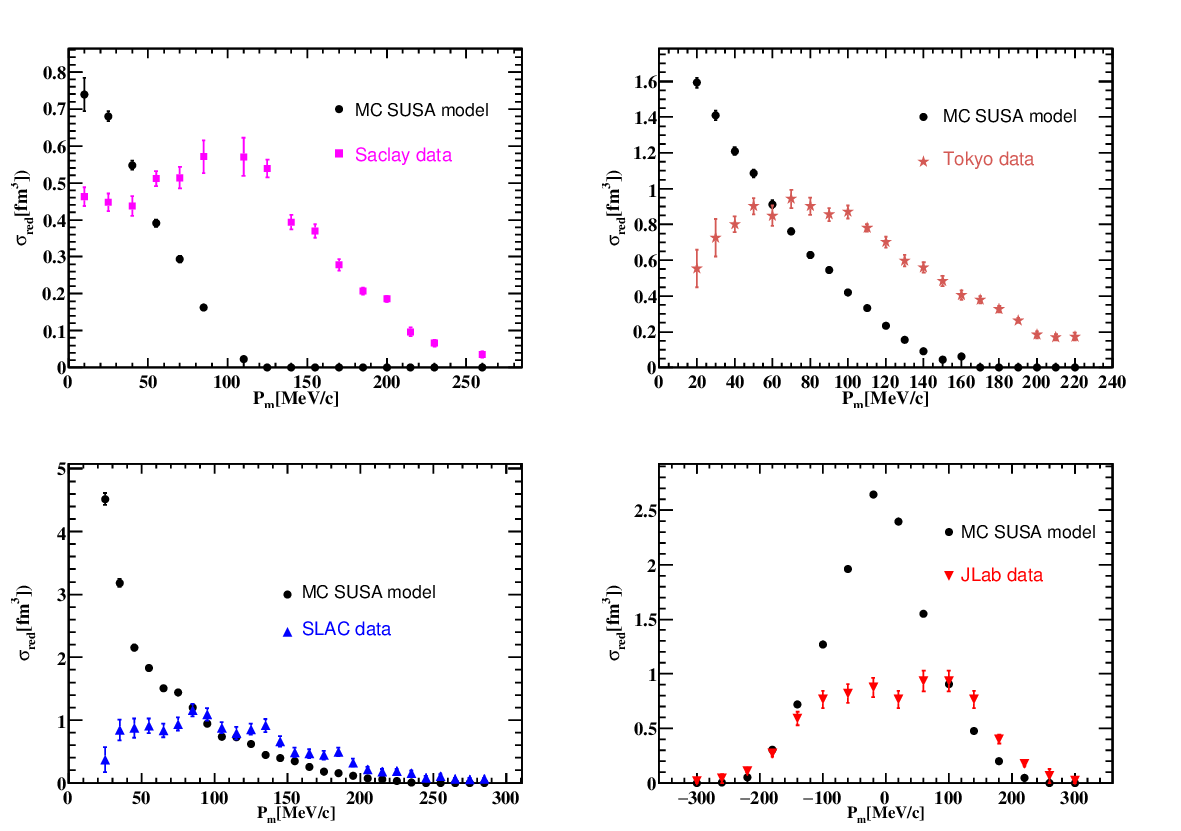}
  \end{center}
  \caption{\label{Fig5}
 Same as Fig.~\ref{Fig3} but for the G21\_11 (SUSAv2) model calculation.    
}
\end{figure*}
The difference between the electron and neutrino reduced cross sections about
10\% can be attributed to Coulomb distortion upon electron wave function.
The electron effective momentum is larger than the value of electron momentum
at infinity, because of Coulomb attraction. The flux is also
increased in the interaction zone by the focusing of the electron wave.
This effect weakens as the beam energy increases, and for this reason this
effect is more significant at Saclay and Tokyo kinematics ($\var_{\nu}=500$ MeV)
than at JLab kinematics ($\var_{\nu}=2.5$ GeV).

The reduced cross section for removing a nucleon in ${}^{12}$C$(\nu_{\mu},\mu p)$
reaction, calculated with the GENIE G18\_10a model, is shown in
Fig.~\ref{Fig4} as
a function of missing momentum together with data. This model uses the LFG
nucleon momentum distribution taking into account the RPA corrections.
\begin{figure*}
  \begin{center}
    \includegraphics[height=12cm, width=18cm]{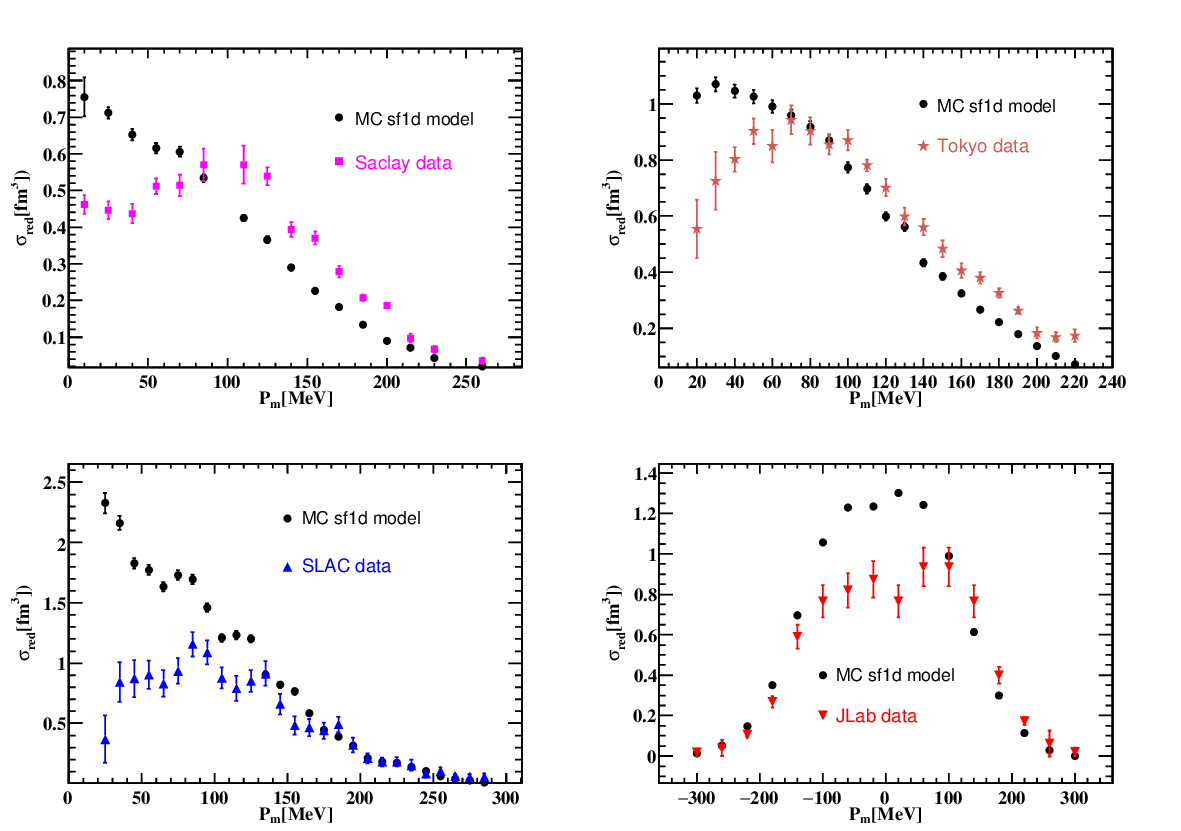}
  \end{center}
  \caption{\label{Fig6}
 Same as Fig.~\ref{Fig3} but for the sf1d (SF) model calculation.    
}
\end{figure*}
At $|p_m| \le 50$ MeV/c the calculated cross sections are significantly higher
than measured ones. The range $|p_m| \le 100$ MeV/c   corresponds to the
proton scattering angles $43^{\circ} \le \theta_p \le 63^{\circ}$. 
The difference decreases as the $p_m$ increases. In the range
$|p_m| \ge 100$ MeV/c there is an overall agreement between calculated and
measured reduced cross sections.

Figure~\ref{Fig5} shows the same as Fig.~\ref{Fig3} but for reduced cross
sections calculated within the G21\_11 model, where LFG nucleon momentum
distribution is
used as well. The $\sigma_{red}$ calculated with this model for Saclay~\cite
{Saclay} and Tokyo~\cite{Tokyo1, Kelly2} kinematics demonstrate absolutely
different behavior than measured ones. At the SLAC~\cite{SLAC} and
JLab~\cite{JLab} kinematics the issues visible at momenta less then 100 MeV/c
resemble those for the G18\_10a model seen in Fig.~\ref{Fig4}.

In Fig.~\ref{Fig6} we compare the GENIE sf1d model calculations with
the data for reduced cross sections of electron scattering on carbon.
It's obvious that the calculated $\sigma_{red}$ at $p_m \le 50$ MeV/c
overestimate the measured ones.
The excess increases with the decrease of missing momentum and reaches up to
50-60\% and higher for the SLAC kinematics. At $p_m \ge 50$ MeV/c the measured
cross sections are lower than calculated $\sigma_{red}$ and discrepancy reduces
with the beam energy ($Q^2$).
\begin{figure*}
  \begin{center}
    \includegraphics[height=12cm, width=18cm]{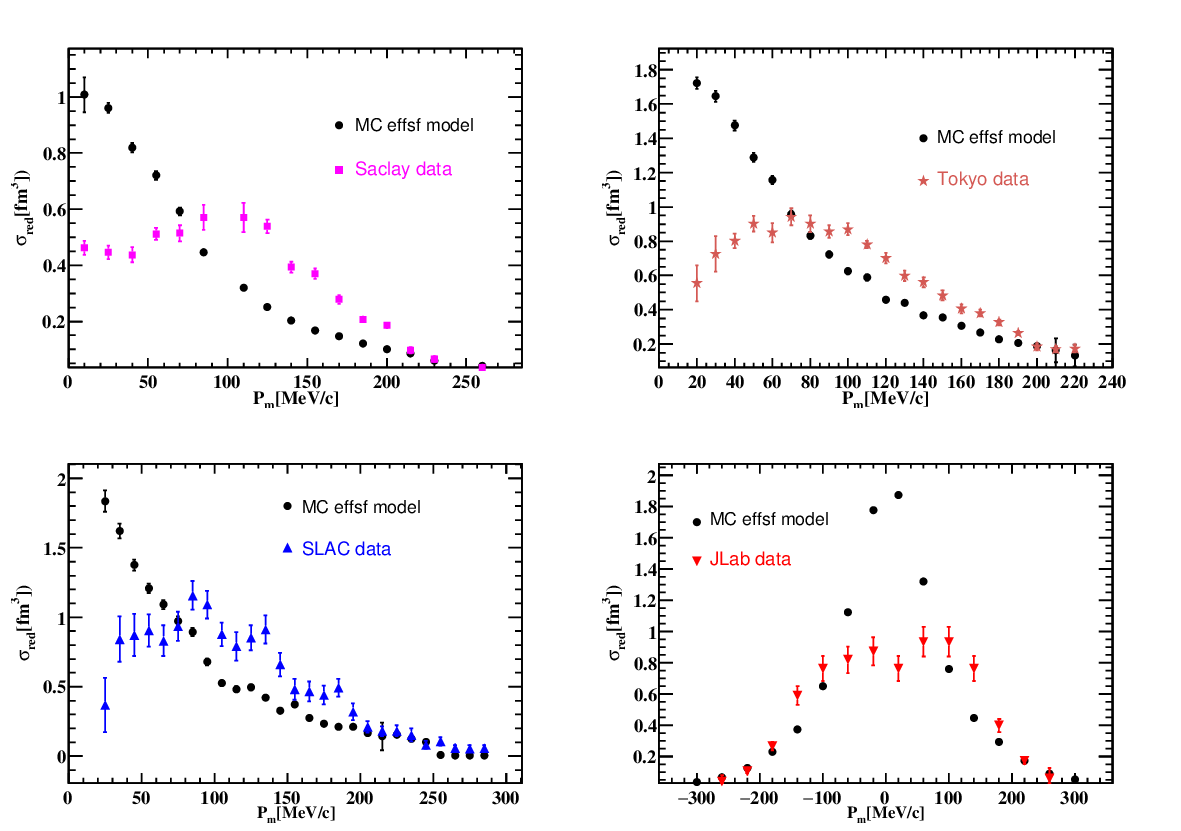}
  \end{center}
  \caption{\label{Fig7}
 Same as Fig.~\ref{Fig3} but for the sf1d (SF) model calculation.    
}
\end{figure*}
The calculations are in a good agreement with
 SLAC and JLab data in the region $p_m \ge 100$ MeV/c.
Approximately the same features are observed in Fig.~\ref{Fig7} where the
comparison with data of the reduced cross sections calculated with the GENIE
effsf model is presented. The calculated cross sections overestimate the
measured ones at $p_m \le 80$ MeV/c and underestimate data at higher missing
momentum. 

So, we observe persistent disagreement between the GENIE predictions of the
reduced cross sections and electron scattering data at low beam energy and
$Q^2$. For the models considered (except the RFG model) the issues are similar:
the calculations overestimate data significantly at low missing momenta and
underestimate gently at $p_m \ge 100$ MeV/c. For example, the best values of
$\chi^2/DOF \approx 12.6$ are observed for the effsf model calculation and 
SLAC data, and $\chi^2/DOF \approx 17.5$ for the sf1d model prediction and 
JLab data. 
\begin{figure*}
  \begin{center}
    \includegraphics[height=12cm, width=18cm]{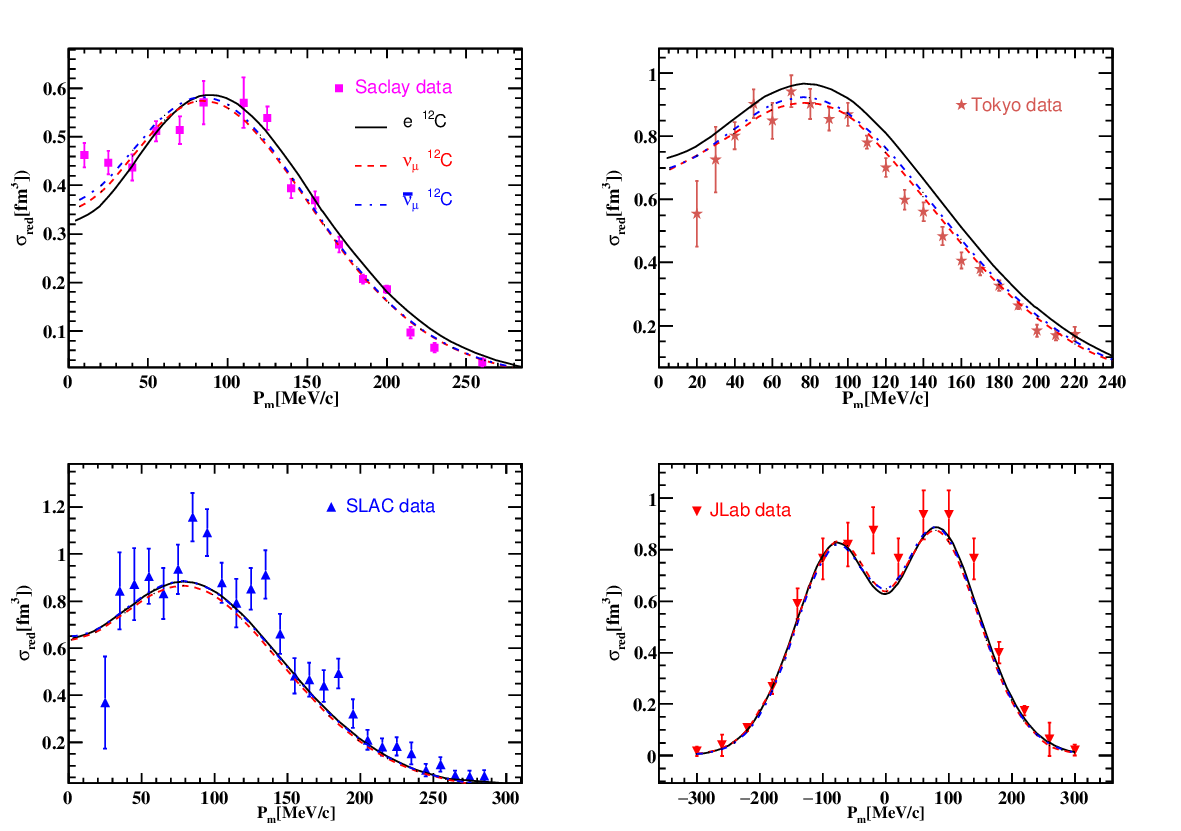}
  \end{center}
  \caption{\label{Fig8}
 Same as Fig.~\ref{Fig3} but for the RDWIA model calculation.    
}
\end{figure*}

The reduced cross sections calculated in Refs.~\cite{BAV2, BAV3, BAV4} within
the relativistic distorted wave impulse approximation (RDWIA) with the EDAD1
parametrization of the optical potential are shown in Fig.~\ref{Fig8} together
with Saclay~\cite{Saclay}, Tokyo~\cite{Tokyo1, Kelly2}, SLAC~\cite{SLAC}, and
JLab~\cite{JLab} data. The cross sections
were calculated for electron and (anti)neutrino scattering on carbon. There is
a good agreement between the RDWIA calculated and measured cross sections. So,
microscopic and unfactorized model like the RDWIA can be used to model both
lepton-boson and boson-nucleus vertexes~\cite{NEUT}.
\begin{figure*}
  \begin{center}
    \includegraphics[height=7cm, width=14cm]{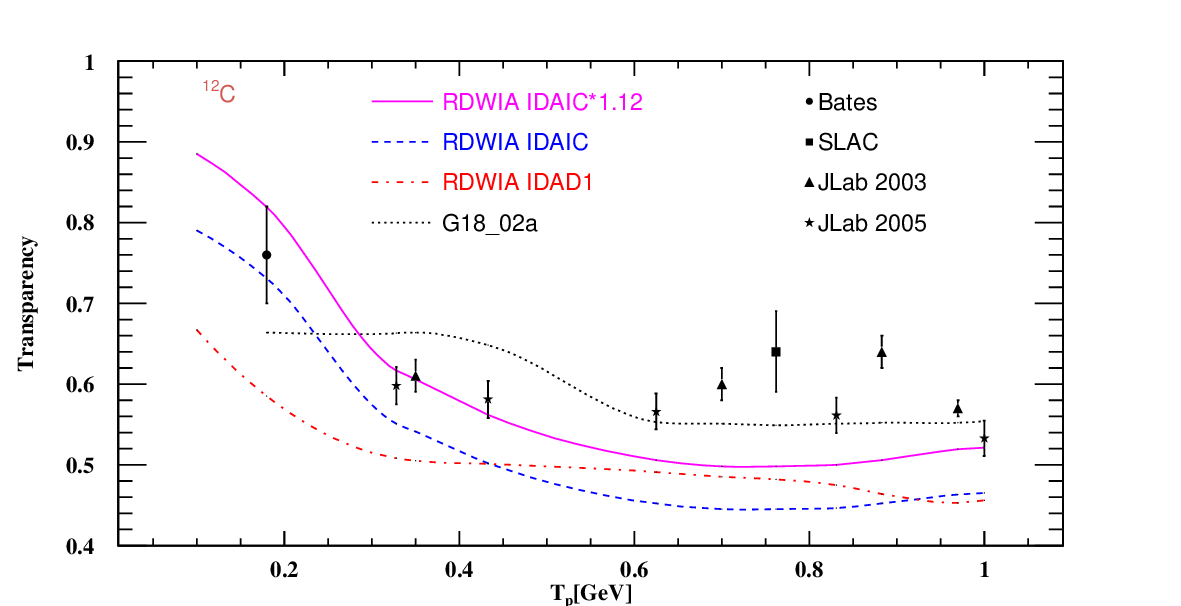}
  \end{center}
  \caption{\label{Fig9}
    Nuclear transparency as a function of proton kinetic energy. Calculations
    compared with measured nuclear transparensy data of Bates~\cite{Bates},
    SLAC~\cite{SLAC}, JLab 2003~\cite{JLab}, and JLab 2005~\cite{JLab05}
    experiments. As shown in the key nuclear transparencies were calculated in
    the RDWIA with the IDAIC, IDAD1 optical potentials, and with GENIE G18\_02a
    model.
}
\end{figure*}
Apparently, this
model describes better the semiexclusive $(l,l'p)$ lepton scattering process
than the phenomenological models employed in the GENIE simulation framework.
In Fig.~\ref{Fig9} the GENIE G18\_02a nuclear  transparency result for carbon
is shown with data for ${}^{12}$C$(e,e'p)$ from Bates~\cite{Bates},
SLAC~\cite{SLAC},
and JLab~\cite{JLab, JLab05}. The transparency curve has a characteristic
shape: a saturation at larger values of proton kinetic energy and decline in
the region of 1 GeV. Saturation can be explained by a roughly constant value
of the total nucleon-nucleon cross sections for larger values of the
incident nucleon momentum. This simulation slightly underestimates data in
the saturation region, but in general the agreement with data is satisfactory.

Also shown in Fig.~\ref{Fig9} are the results obtained by Kelly in the RDWIA
using several parametrizations of the optical potential~\cite{Kelly2}. All of
these calculations systematically underestimate the experimental transparency.
Multiplying the EDAIC results by 1.12 provides a reasonable description of the
data over this range of ejectile energies. The author concluded that a more
detailed model of the continum arising from multinucleon knockout that
contribute approximately 12\% of strength for $\var_m \le 80$ MeV is needed.

\begin{figure*}
  \begin{center}
    \includegraphics[height=8cm, width=16cm]{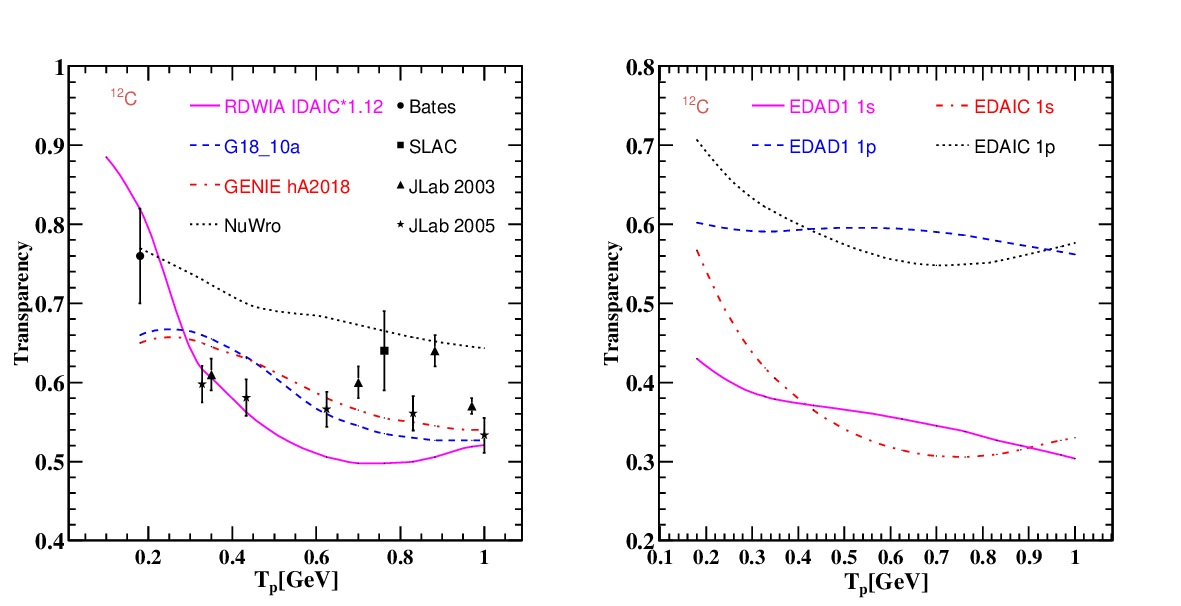}
  \end{center}
  \caption{\label{Fig10}
    Nuclear transparency as a function of proton kinetic energy. Left panel:
    the RDWIA calculation with the IDAIC optical potential, G18\_10a model
    and NuWro generator results compared with measured nuclear transparency
    data of Bates~\cite{Bates}, SLAC~\cite{SLAC}, JLab 2003~\cite{JLab}, and
    JLab 2005~\cite{JLab05} experiments. Right panel: transparency for the
    $1p_{3/2}$ and $1p_{1/2}$ orbits in ${}^{12}$C as obtained in the RDWIA with
    the IDAIC and IDAD1 optical potentials. 
}
\end{figure*}

Comparison of the EDADIC result multiplied by 1.12 and result of the calculation
with the G18\_10a model performed in this work is shown in Fig.~\ref{Fig10} on
the left panel. Also shown are the results obtained in Ref.~\cite{Dytman} with
the GENIE G18\_10a model (GENIE hA2018) and in Ref.~\cite{NuWro3} using the
NuWro neutrino event generator.
The results obtained within the G18\_10a in this work and in
Ref.~\cite{Dytman} are in good agreement. On the other hand the values of
the NuWro nuclear transparency are systematically larger than those obtained
in our calculation. In Fig.~\ref{Fig10} on the right panel the nuclear
attenuations calculated within the RDWIA for the individual shells in ${}^{12}$C
are presented. The calculations were performed with the EDAD1 and EDAIC
approximations of the optical potential. Investigating the attenuation for
each individual shell allow one to study the radial dependence of the FSI
mechanisms. In the ${}^{12}$C case, for example, the $1s_{1/2}$ orbit has spatial
characteristics which are very different from the $1p_{3/2}$ orbit. As expected,
the RDWIA model predict a stronger attenuation for proton emission from a level
which has a larger fraction of its density in the nuclear interior.
\section{Conclusions}
In this paper we show that reduced cross sections and nuclear tansparencies
measured in QE electron-scattering experiments can serve as a very effective
tool for testing neutrino event generators. This is made possible by two
factors. First these experiments features precisely known electron and proton
kinematics:
initial electron energy and energies of the scattered electron and proton as
well as their scattering angles. Second there are numerous (at least for
carbon) electron-scattering reduced cross section datasets covering the range
of energies and angles relevant to neutrino oscillation experiments.
We carried out a systematic comparison of the CCQE reduced
cross sections and nuclear transparency calculated with the models employed
in the GENIE v3 simulation framework with the data measured in electron
scattering off carbon target. Because these data have not been previously used
to tune the parameters of the neutrino events generators they present great
opportunity to test the accuracy of the cross section estimates.

We have observed persistent disagreements
between the GENIE predictions and electron scattering data for reduced cross
sections. Below missing momentum $p_m\approx 80$ MeV/c the reduced cross
sections predicted by the GENIE models based on the LFG nucleon momentum
distribution (G18\_10a and G21\_11 models) overestimate data significantly.
On the other hand apparently that microscopic and unfactorized model like the
RDWIA describes the semiexclusive lepton scattering reactions better than
the phenomenological models employed in the GENIE generator.

The study of the nuclear transparency presented in this paper showed that
agreement of the GENIE calculations with data is satisfactory. But, we found
that the GENIE models as well as the RDWIA should be enriched with
additional effects that are not directly related to the single-nucleon
knockout. So, the direct comparison of spectral functions,
implemented in neutrino event generators, with the precise electron reduced
cross section data presents a great opportunity to test better the accuracy of
the nuclear effect calculations.

\section*{Acknowledgments}

The author greatly acknowledge A. Habig for fruitful discussions and a
critical reading of the manuscript. 
%



\begin{thebibliography}{99}
\bibitem{NOvA1} M.~A.~Acero {\it et al.}, (NOvA Collaboration), Phys. Rev. Lett.
{\bf 123}, 151803 (2019).
\bibitem{T2K} K.~Abe {\it et al.}, (T2K Collaboration), Phys. Rev. Lett.
{\bf 121}, 171802 (2018).
\bibitem{DUNE} R.~Acciarri {\it et al.}, (DUNE Collaboration), 
  FERMILAB-DESIGN-2016-03.
\bibitem{HK2T} K.~Abe {\it et al.}, (Hyper-Kamiokande Collaboration)
  arXiv:1805.04163 [physics.ins-det].
\bibitem{SBN} M.~Antonello {\it et al.} (MicroBooNE, LAr1-ND, ICARUS-WA104
  Collaboration), arXiv:1503.01520 [physics.ins-det].
\bibitem{Pick}  A.~Picklesimer, J.~W.~Van Orden, S.~J.~Wallace,
Phys. Rev. {\bf C 32}, 1312 (1985).
\bibitem{Udias} J.~M.~Udias, P.~Sarriguren, E.~Moya de Guerra, E.~Garrido, and
J.~A.~Caballero, Phys. Rev. {\bf C 51}, 3246 (1995).
\bibitem{JKelly} James~J.~Kelly, Phys. Rev. {\bf C 59}, 3256 (1999).
\bibitem{Gon3} R.~Gonzalez-Jimenez, M.~B.~Barbaro, J.~A.~Caballero,
  T.~W.~Donnelly, N.~Jachowicz, G.~D.~Megias, K.~Niewczas, A.~Nikolakopoulos,
  J.~M.~Udias, Phys. Rev. {\bf C 101}, 015503 (2020).
\bibitem{BAV7} A.~V.~Butkevich and S.~V.~Luchuk, Phys. Rev. {\bf C 102} 024602
  (2020).
\bibitem{Megias2} G.~D.~Megias, J.~E.~Amaro, M.~B.~Barbaro, J.~A.~Caballero, 
T.~W.~Donnelly, Phys. Rev. {\bf D 94}, 013012 (2016).
\bibitem{Megias3} G.~D.~Megias, J.~E.~Amaro, M.~B.~Barbaro, J.~A.~Caballero, 
T.~W.~Donnelly, and I.~R.~Simo, Phys. Rev. {\bf D 94}, 093004 (2016).  
\bibitem{Ankow}  A.~M.~Ankowski and A.~Friedland, Phys. Rev. {\bf D 102}, 053001
  (2020).
\bibitem{e4v1}  A.~Papadopoulou {\it et al.} (e4v Collaboration)
 Phys. Rev. {\bf D 103}, 113003 (2021).
\bibitem{CLAS} M.~Khachatryan, A.~Papapdopoulou, A.~Ashkenazi, F.~Hauenstein,
  L.~B.~Weinstein, O.~Hen, E.~Piasetzky  (CLAS and e4v Collaboration), Nature
  {\bf 599}, 565 (2021).
\bibitem{Ruso} L.~Avarez-Ruso, {\it et al.} (GENIE Collaboration), Eur. Phys.
 J. Spec. Top. {\bf 230}, 4449 (2021).
\bibitem{NuWro1} J.~Zmuda, K.~Graczyk, C.~Juszczak, and J.~Sobczyk, 
 Acta. Phys. Pol. B46, 2329 (2015).
\bibitem{NuWro2} R.~D.~Banerjee, A.~M.~Ankowski, K.~M.~Graczyk, B.~E.~Kowal,
  H.~Prasad, and J.~T.~Sobczyk, Phys. Rev. {\bf D 109}, 073004 (2024).
\bibitem{ACHI} J.~Isaacson, W.~J.~Jay, A.~Lovato, P.~A.~N.~Machado, and
  N.~Rocco, Phys. Rev. {\bf D 107}, 033007 (2023).
\bibitem{NEUT} S.~Abe, Phys. Rev. {\bf D 111}, 033006 (2025).
\bibitem{Dytman} S.~Dytman, Y.~Hayato, R.~Raboanary, J.~T.~Sobczyk,
  J.~Tena Vidal and N.~Vololoniaina, Phys. Rev. {\bf D 104}, 053006 (2021).  
\bibitem{NuWro3} K.~Niewczas, and J.~T.~Sobczyk, Phys. Rev. {\bf C 100},
  015505 (2019).
\bibitem{BAV1} A.~V.~Butkevich and S.~A.~Kulagin, Phys. Rev. {\bf C 76},
  045502 (2007).
\bibitem{BAV2} A.~V.~Butkevich, Phys. Rev. {\bf C 80}, 014610 (2009).
\bibitem{BAV3} A.~V.~Butkevich, Phys. Rev. {\bf C 85}, 065501 (2012).
\bibitem{BAV4} A.~V.~Butkevich, Phys. Rev. {\bf C 109}, 045502 (2024).
\bibitem{GEN1} C.~Andreopoulos, A.~Bell, D.~Bhattacharya, F.~Cavanna,
J.~Dobson, S.~Dytman, H.~Gallagher, P.~Guzowski, R.~Hatcher, P.~Kehayias
 {\it et al}, Nucl. Instrum. Meth. {\bf A614}, 87 (2010).
\bibitem{GEN2} The GENIE Event Generator, http://genie-mc.org 
\bibitem{Caballero} J.~A.~Caballero, T.~W.~Donnelly, E.~Moya~de~Guerra, and
  J.~M.~Udias, Nucl. Phys. {\bf A 632}, 323 (1998).
\bibitem{Sanchez} A.~Nikolakopoulos, R.~Gonzalez-Jimenez, N.~Jachowicz,
  K.~ Niewczas, F.~Sanchez, J.~M.~Udias, Phys. Rev. {\bf C 105} 054603 (2022).  
\bibitem{Tokyo1} K.~Nakamura, and N.~Izutsu, Nucl. Phys. {\bf A 259}, 301
  (1976).
\bibitem{Kelly2} J.~J.~Kelly, Phys. Rev. {\bf C 71}, 064610 (2005).
\bibitem{Saclay} J.~Mougey, M.~Bernheim, A.~Bussiere, A.~Gillebert,
  P.~X.~Ho, M.~Priou, D.~Royer,I.~Sick, and G.~Wagner, Nucl. Phys. {\bf A 262},
  461 (1976).
\bibitem{SLAC}  
  N.~C.~R.~Makins, R.~Ent, M.~S.~Chapman, J.~O.~Hansen, K.~Lee, Milner~R.G.
  {\it et al.}, Phys. Rev. Lett. {\bf 72}, 1986 (1994).
\bibitem{JLab} D.~Dutta {\it et al.}, Phys. Rev. {\bf C 68}, 064603 (2003).
\bibitem{G1} C.~Andreopoulos, C.~Barry, S.~Dytman, H.~Gallagher, T.~Golan,
  R.~Hatcher, G.~Perdue, J.~Yarba, arXiv:1510.05494.
\bibitem{G2} L.~Alvarez-Russo {\it et al.}, (GENIE Collaboration), Eur. Phys.
  J. Spec. Top. {\bf 230}, 4449 (2021).
\bibitem{G3} J.~Tena-Vidal {\it et al.}, (GENIE Collaboration), Phys. Rev.
  {\bf D 106}, 112001 (2022).  
\bibitem{GEM} S.~J.~Mashnik, A.~J.~Sierk, K.~K.~Gudima, and M.~I.~Baznat, J.
 Phys. Conf. Ser. {\bf 41}, 340 (2006).
\bibitem{BodR} A.~Bodek and J.~L.~Ritchie, Phys. Rev. {\bf D 23}, 1070 (1981).
\bibitem{LS} C.~Llewellyn~Smith, Phys. Rep. {\bf 3}, 261 (1972).
\bibitem{Val1} J.~Nieves, I.~Ruiz~Simo, and M.~J.~Vicente~Vacas, Phys. Lett. 
{\bf B 707}, 72 (2012).
\bibitem{Val2} J.~Nieves, J.~E.~Amaro, and M.~Valverde, Phys. Rev. {\bf C 70},
  055503 (2004).Phys. Rev. D106, 112001 (2022)
\bibitem{Val3} R.~Gran, J.~Nieves, F.~Sanchez, and M.~J.~Vicente~Vacas,
  Phys. Rev. {\bf D88}, 113007 (2013). 
\bibitem{Dolan} S.~Dolan, G.~D.~Megias, and S.~Bolognesi, Phys. Rev.
  {\bf C 101}, 033003 (2020).
\bibitem{Benh1} O.~Benhar, A.~Fabrocini, S.~Fantoni, Nucl. Phys. {\bf A 505},
  267 (1989).
\bibitem{Benh2} O.~Benhar, A.~Fabrocini, S.~Fantoni, Nucl. Phys. {\bf A 579},
  493 (1994).
\bibitem{Bod1} A.~Bodek, H.~S.~Budd, M.~E.~Christy, Eur. Phys. J. {\bf C 71},
 1726 (2011).
\bibitem{Bod2} A.~Bodek, H.~S.~Budd, M.~E.~Christy, B.~Coopersmith, Eur. Phys.
  J. {\bf C 74}, 3091 (2014).
\bibitem{NEUT} J.~McKean, R.~Gonzalez-Jimenez, M.~Kabirnezhad, J.~M.~Udias,
  arXiv:2502.10629
\bibitem{Bates} G.~Garino, {\it et al.}, Phys. Rev. {\bf C 45}, 780  (1992).
\bibitem{JLab05} D.~Rohe {\it et al.}, Phys. Rev. {\bf C 72}, 054602 (2005).

\end{thebibliography}
\end{document}